\documentclass[aps,prl,preprint,superscriptaddress,showpacs]{revtex4-1}

\usepackage{graphicx}
\usepackage{dcolumn}
\usepackage{bm}
\usepackage{amsmath}
\usepackage{amssymb}
\usepackage{latexsym}
\usepackage{epsfig}
\usepackage{amsbsy}
\usepackage{array}
\usepackage{amssymb}
\usepackage{setspace}
\usepackage{bm}

\begin{document}


\title{NMR and NQR studies on transition-metal arsenide superconductors LaRu$_2$As$_2$, KCa$_2$Fe$_4$As$_4$F$_2$, and A$_2$Cr$_3$As$_3$}
\thanks{These authors contribute equally.}

\author{Jun Luo*}
\email[]{junluo@iphy.ac.cn.}
\affiliation{Institute of Physics, Chinese Academy of Sciences,\\
 and Beijing National Laboratory for Condensed Matter Physics, Beijing 100190, China}

\author{Chunguang Wang*}
\affiliation{Institute of Physics, Chinese Academy of Sciences,\\
 and Beijing National Laboratory for Condensed Matter Physics, Beijing 100190, China}
\affiliation{School of Physical Sciences, University of Chinese Academy of Sciences, Beijing 100049, China}

\author{Zhicheng Wang}
\affiliation{Department of Physics, Zhejiang University, Hangzhou 310027, China}

\author{Qi Guo}
\affiliation{Institute of Physics, Chinese Academy of Sciences,\\
 and Beijing National Laboratory for Condensed Matter Physics, Beijing 100190, China}

\author{Zhicheng Wang}
\affiliation{Department of Physics, Zhejiang University, Hangzhou 310027, China}

\author{Jie Yang}
\affiliation{Institute of Physics, Chinese Academy of Sciences,\\
 and Beijing National Laboratory for Condensed Matter Physics, Beijing 100190, China}

\author{Rui Zhou}
\affiliation{Institute of Physics, Chinese Academy of Sciences,\\
 and Beijing National Laboratory for Condensed Matter Physics, Beijing 100190, China}
\affiliation{Songshan Lake Materials Laboratory, Dongguan, Guangdong 523808, China}

\author{K. Matano}
\affiliation{Department of Physics, Okayama University, Okayama 700-8530, Japan}

\author{T. Oguchi}
\affiliation{Institute of Scientific and Industrial Research, Osaka University, Osaka
567-0047, Japan}

\author{Zhian Ren}
\affiliation{Institute of Physics, Chinese Academy of Sciences,\\
 and Beijing National Laboratory for Condensed Matter Physics, Beijing 100190, China}
 \affiliation{School of Physical Sciences, University of Chinese Academy of Sciences, Beijing 100049, China}

\author{Guanghan Cao}
\affiliation{Department of Physics, Zhejiang University, Hangzhou 310027, China}

\author{Guo-qing Zheng }
\email[]{gqzheng123@gmail.com}
\affiliation{Institute of Physics, Chinese Academy of Sciences,\\
 and Beijing National Laboratory for Condensed Matter Physics, Beijing 100190, China}
 \affiliation{Department of Physics, Okayama University, Okayama 700-8530, Japan}

\date{\today}

\begin{abstract}
We report $^{75}$As-nuclear magnetic resonance (NMR) and nuclear quadrupole resonance (NQR) measurements on transition-metal arsenides LaRu$_2$As$_2$, KCa$_2$Fe$_4$As$_4$F$_2$, and A$_2$Cr$_3$As$_3$. In the superconducting state of LaRu$_2$As$_2$, a Hebel-Slichter coherence peak is found in the temperature dependence of the spin-lattice relaxation rate-1/$T_1$ just below $T_{\rm c}$, which indicates that LaRu$_2$As$_2$ is a full-gap superperconducor. For KCa$_2$Fe$_4$As$_4$F$_2$, 
antiferromagnetic spin fluctuations are observed in the normal state. We further find that the anisotropy rate $R_{\rm AF}$ = $T_{1}^{c}$/$T_{1}^{ab}$ is small and temperature independent, implying that the low energy spin fluctuations are isotropic in spin space. Our results indicate that KCa$_2$Fe$_4$As$_4$F$_2$ is a moderately overdoped iron-arsenide high-temperature superconductor with a stoichiometric composition. For A$_2$Cr$_3$As$_3$, we calculate the electric field gradient by first-principle method and assign the $^{75}$As-NQR peaks with two
crystallographically different As sites, paving the way for further NMR investigation.
\end{abstract}


\maketitle

\section{}
Transition metal arsenides (TMAs) belong to a big family. The binary TMAs, like TaAs, TaP, NbP and XP$_2$ (X= Mo, W)~\cite{WHM}, are topological Weyl semimetals, whose low energy excitations in the bulk can be viewed as chiral massless Weyl Fermions. The ternary TMAs show rich novel properties, with examples including density wave~\cite{densitywave} and superconductivity~\cite{122superconductivity}. The discovery of superconductivity in transition metal arsenide LaFeAsO$_{1-x}$F$_x$ opens a door to another high-temperature superconducting family besides cuprates~\cite{LaFeAsOF}. More importantly, the physical properties of TMAs can be tuned by chemical substitution~\cite{AFe2As2,AFe2As2Kondo}, doping~\cite{LaFeAsOF}, or pressure~\cite{CrAs}. Therefore, transition-metal arsenides provide a rich material base for exploring exotic physical phenomena.

Among the TMA family, compounds with ThCr$_2$Si$_2$-type layered crystal structure has attracted many attentions in condensed matter physics. High temperature superconductivity in AFe$_2$As$_2$ (A = Ca, Sr, Ba, etc.) was induced by doping~\cite{BaFeAsdoping} or pressure~\cite{BaFeAspressure}. Iron and ruthenium are in the same group. Many Ru-based compounds show unconventional superconductivity~\cite{SrRuO,URuSi}. Naturally, it is practical to look for unconventional superconductivity in ruthenium-based compounds with ThCr$_2$Si$_2$ structure. Recently, Guo $et$ $al$ found that LaRu$_2$As$_2$ shows superconductivity with zero resistivity at 6.8 K~\cite{LaRuAs}. 
LaRu$_2$As$_2$ and LaRu$_2$P$_2$ are isostructural and their physical properties have been studied by $ab$ $initio$ calculations~\cite{LaRuAsab,LaRuAsDFT}, which indicate that the conduction band electrons are mainly contributed from La-5$d$ and Ru-4$d$ orbitals. 
KCa$_2$Fe$_4$As$_4$F$_2$ is a newly discovered superconductor with separated double Fe$_2$As$_2$ layer, whose $T_{\rm c}$ reaches as high as 33.5 K~\cite{KCa2Fe4As4F2}. It can be regarded as the intergrowth of 1111-type CaFeAsF and 122-type KFe$_2$As$_2$~\cite{KCa2Fe4As4F2}. 
The Fe valence of CaFeAsF and KFe$_2$As$_2$ are +2 and +2.5, respectively. Appointing the insulted compound CaFeAsF as the parent compound, KCa$_2$Fe$_4$As$_4$F$_2$ can be viewed as a self hole-doping system, which is consistent with the Hall effect measurements and electronic structures calculations~\cite{KCa2Fe4As4F2,electronicstructureKCa2Fe4As4F2}. 
In the superconducting state, inverse square penetration depth ($\lambda_{ab}^{-2}$) with a linear temperature dependence detected by muon spin rotation ($\mu$SR) suggests line node in the gap function in KCa$_2$Fe$_4$As$_4$F$_2$ and CsCa$_2$Fe$_4$As$_4$F$_2$~\cite{uSRCsCa2Fe4As4F2,uSRKCa2Fe4As4F2}. However, optical conductivity, thermal conductivity  and ARPES measurements suggest a nodeless gap~\cite{opticsCsCaFeAsF,heatCsCaFeAsF,ARPESKCaFeAsF}.

A$_2$Cr$_3$As$_3$ (A = Na, K, Rb, Cs) is the first chromium-based superconducting family under ambient pressure, with $T_{\rm c}$ ranging from 8.0K to 2.2K\cite{K2Cr3As3,Rb2Cr3As3,Cs2Cr3As3,Na2Cr3As3}. In the crystal structure, [Cr$_{3}$As$_{3}$]$_\infty$ chains are separated by alkaline metal. Owing to asymmetric distribution of alkaline metal, there exists two types of As sites. Density function theory (DFT) calculations show that the fermi surface is formed by one three-dimension (3D) band $\gamma$ and two quasi-one-dimension (1D) band $\alpha$ and $\beta$\cite{ferromagneticfluctuation1,magnetism}. Experimental results point to unconventional superconductivity in A$_2$Cr$_3$As$_3$~\cite{imai,yang,penetration,muSRK2Cr3As3,muSRRb2Cr3As3,
angularHc2,specific}. In the normal state, ferromagnetic fluctuation (FM) was revealed~\cite{yang} and further found to be enhanced by small radius alkaline metal replacement~\cite{FMQCP}. Similar to iron-pnictide superconductors, A$_2$Cr$_3$As$_3$ shows a close relationship between magnetic fluctuations and superconductivity. More recently, nontrivial topological aspects in A$_2$Cr$_3$As$_3$ has been pointed out~\cite{FMQCP,topology}.

In this work, we performed nuclear magnetic resonance(NMR) and nuclear quadrupole resonance(NQR) measurements on LaRu$_2$As$_2$, KCa$_2$Fe$_4$As$_4$F$_2$ and A$_2$Cr$_3$As$_3$. We investigate the properties of LaRu$_2$As$_2$ and KCa$_2$Fe$_4$As$_4$F$_2$. For A$_2$Cr$_3$As$_3$, we assign the one to one correspondence between the two NQR transition lines and the two As crystallographic sites, by combining the first-principle calculation. 

\section{Experiment}  
\noindent Polycrystalline LaRu$_2$As$_2$ and KCa$_2$Fe$_4$As$_4$F$_2$ samples were grown by conventional solid state reaction, as previously reported in Ref.~\cite{LaRuAs,KCa2Fe4As4F2}. The single crystals of A$_2$Cr$_3$As$_3$ were synthesized by high-temperature solution method with A = Cs, Rb , K, Na$_{0.75}$K$_{0.25}$ or ion-exchanged reaction with A = Na, the details of synthesis can be seen in ref~\cite{K2Cr3As3,Rb2Cr3As3,Cs2Cr3As3,Na2Cr3As3}. The $T_{\rm c}$ of LaRu$_2$As$_2$ and KCa$_2$Fe$_4$As$_4$F$_2$ were determined by AC susceptibility using an $in-situ$ NMR coil. The EFG of K$_2$Cr$_3$As$_3$ are calculated by the all electron Full-Potential Linear Augmented Plane Wave (FLAPW) method implemented in Hiroshima Linear-Augmented-Plane-Wave (HiLAPW) code with generalized gradient approximation including spin-orbit coupling~\cite{HILAPW}. The NMR and NQR spectra were obtained by scanning RF frequency and integrating spin-echo intensity at a fixed magnetic field $H_{0}$. The spin-lattice relaxation time $T_{1}$ was measured by the saturation-recovery method. $T_{1}$ was obtained by fitting the nuclear magnetization $M(t)$  to $1-M(t)/M_{0} = \exp(-3t/T_{1})$ in NQR case and $1-M(t)/M_{0} = 0.1\exp(-t/T_{1})+0.9\exp(-6t/T_{1})$ in NMR case, where $M(t)$ and $M_0$ are the nuclear magnetization at time $t$ after the single comb pulse and at thermal equilibrium, respectively.

\section{Results and discussions}  
\subsection{LaRu$_2$As$_2$}
\begin{figure}[htbp]
\includegraphics[width=6 cm]{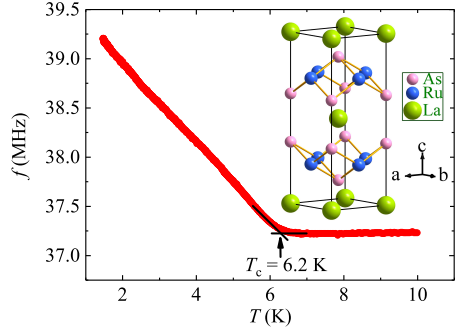}
\centering
\caption{(Color online) Temperature dependence of the NMR coil resonance frequency. $T_{\rm c}$ is determined by the cross point of the two straight lines shown in the figure. The inset shows crystal structure of LaRu$_2$As$_2$.
\label{LaRu2As2susceptibility1}}
\end{figure}

Figure~\ref{LaRu2As2susceptibility1} shows temperature dependence of the resonance frequency of the NMR coil. The superconducting transition temperature $T_{\rm c}$ of the sample is found to be around 6.2 K, which is similar to an earlier report of $T_{\rm c}$ = 6.8K determined by DC susceptibility measurement~\cite{LaRuAs}. There are two primitive cells with ten atoms in one unit cell of LaRu$_2$As$_2$ as shown in the inset of Fig.~\ref{LaRu2As2susceptibility1}. The As-Ru-As blocks are intercalated by La atoms, and they are alternately arranged along the $c$ axis. As a result, all As sites are equivalent in LaRu$_2$As$_2$.

\begin{figure}[htbp]
\includegraphics[width=5 cm]{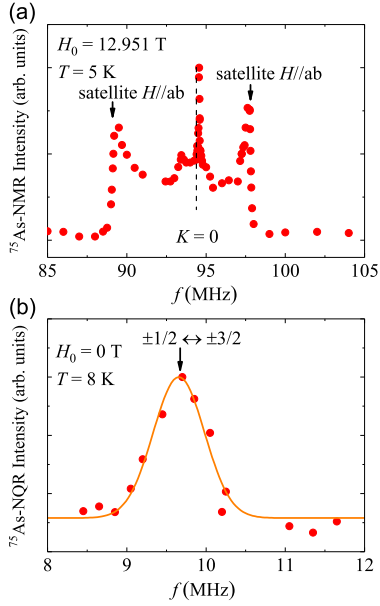}
\centering
\caption{(Color online) (a) $^{75}$As NMR spectrum of LaRu$_2$As$_2$ at $T$ = 5 K and $\mu_0H_0$ = 12.951 T. There are two satellite peaks. The central transition frequency $\nu_{\rm res}$ has two peaks with $\theta$ = 42$^\circ$ and $\theta$ = 90$^\circ$ due to second order perturbation. The dashed line corresponds to $K = 0$. (b)$^{75}$As NQR spectrum of LaRu$_2$As$_2$ at $T$ = 8 K. The curve is Gaussian fitting of the spectrum.
\label{LaRu2As2spec1}}
\end{figure}

Figure~\ref{LaRu2As2spec1}(a) shows the $^{75}$As NMR spectrum measured at $T$ = 5 K under magnetic field of $\mu_0H_0$ = 12.951 T, which is a typical powder pattern for nucleus with spin $I$ = 3/2. Considering the tetragonal lattice of LaRu$_2$As$_2$, the total Hamiltonian for $I$ = 3/2 nucleus can be expressed as~\cite{Abragam}:
\begin{equation}
\mathcal{H} = \mathcal{H}_0 + \mathcal{H}_Q = \gamma\hbar I_zH_0(1+K)+\frac{e^2qQ}{4I(2I-1)}[3(I_z^2-I^2)(3\rm{cos}^2\theta-1)]
\end{equation}
where $K$ is the Knight shift, $eq$ = $V_{ZZ}$ = $\frac{ \partial{V^2}}{\partial {Z^2}}$ is the electric field gradient (EFG) along principle axis $Z$, $Q$ is the nuclear quadrupole moment, $\theta$ is the angle between the magnetic field and the principle axis of the EFG. The nuclear quadrupole resonance frequency $\nu_{\rm Q}$ is defined as $\frac{3e^2qQ}{2I(2I-1)h}$. The two peaks (marked by two black arrows) observed at 89.2 MHz and 97.8 MHz correspond to the transitions (3/2 $\leftrightarrow$ 1/2) and (-1/2 $\leftrightarrow$ -3/2). The central transition frequency $\nu_{\rm res}$ for $I$ = 3/2 to the second order is given by~\cite{Abragam}
\begin{equation}
\nu_{\rm res} = \gamma H_0(1+K) + \frac{3\nu_{\rm Q}^2}{16\gamma H_0(1+K)}\rm {sin}^2\theta(1-9\rm{cos}^2\theta)
\end{equation}
The observed spectrum is in agreement with the theoretically expected characteristic powder pattern with two peaks at $\theta$ = 42$^\circ$ and $\theta$ = 90$^\circ$. For
a randomly oriented powder sample, the peak at $\theta$ = 42$^\circ$ would have a larger intensity than that at $\theta$ = 90$^\circ$. Thus our observation suggests that the powder is partially oriented. Fig.~\ref{LaRu2As2spec1}(b) shows the NQR spectrum, in which only one transition ($\pm1/2 \leftrightarrow \pm3/2$) is observed, considering that there is only one As site in this compound. We fitted the $^{75}$As NQR spectrum by a Gaussian function, and deduced $\nu_{\rm Q}$ = 9.65 MHz.

\begin{figure}[htbp]
\includegraphics[width=6 cm]{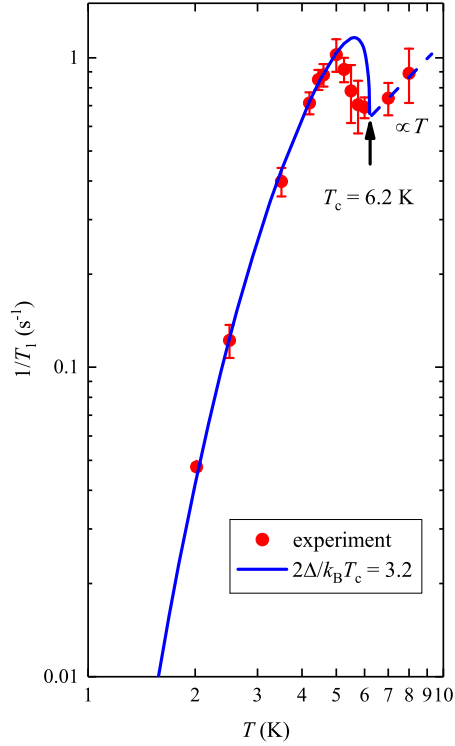}
\centering
\caption{(Color online) The temperature dependence of $^{75}$As NQR spin-lattice relaxation rate $1/T_1$. $1/T_1$ shows a coherence peak just below $T_{\rm c}$. The dashed line shows a $T_1T$ = constant relation. The solid curve is the fitting result assuming an s-wave gap.
\label{LaRu2As2T1}}
\end{figure}
Figure~\ref{LaRu2As2T1} shows the temperature dependence of $1/T_1$ measured via $^{75}$As
NQR. The blue dashed line is a guide to the eyes showing the relation of $T_1T$ = constant. We note that $1/T_1$ shows a clear Hebel-slichter peak just below $T_{\rm c}$ = 6.2 K and decreases exponentially at low temperatures, which are characteristics of an isotropic superconducting gap. The relaxation rate below $T_{\rm c}$ is expressed as~\cite{Maclaughlin}
\begin{equation}
\frac{T_1(T_{\rm c})}{T_{\rm 1s}} = \frac{2}{k_BT_{\rm c}}\int N_{\rm s}(E)^2(1+\frac{\Delta^2}{E^2})f(E)(1-f(E))dE
\end{equation}
where $\Delta$ is the magnitude of energy gap, $N_{\rm s}(E)$ = $N_0\frac{E}{\sqrt{E^2-\Delta^2}}$ is the DOS in superconducting state, $(1+\frac{\Delta^2}{E^2})$ is the coherence factor, and $f(E)$ is Fermi distribution function. We convolute $N_{\rm s}(E)$ to a rectangular broadening function with a width $2\delta$ and a height $1/2\delta$~\cite{Hebel}. The solid curve in Fig.~\ref{LaRu2As2T1} is the simulation with the parameters 2$\Delta/k_{\rm B}T_{\rm c}$ = 3.2 and $r$ = $\Delta/\delta$ = 12, which is in good agreement with the experimental data. The value of 2$\Delta/k_{\rm B}T_{\rm c}$ is close to the BCS value of 3.5.


\subsection{KCa$_2$Fe$_4$As$_4$F$_2$}
\begin{figure}[htbp]
\includegraphics[width= 10 cm]{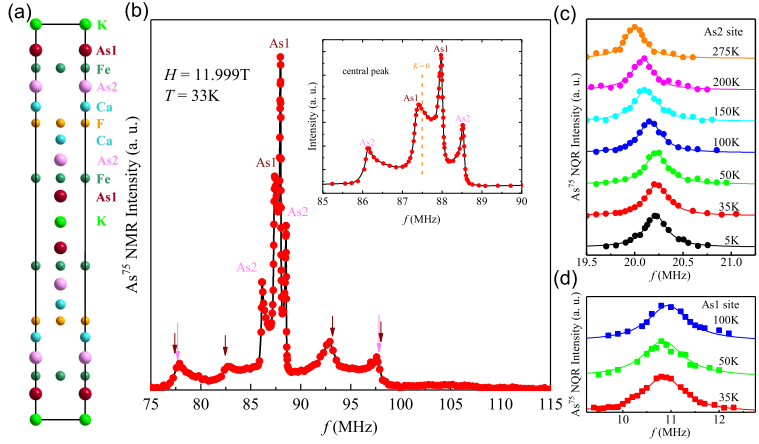}
\centering
\caption{(Color online) (a)Crystal structure of KCa$_2$Fe$_4$As$_4$F$_2$. There are two type of As sites marked as dark red color and pink color. (b)Nuclear magnetic resonance spectrum of KCa$_2$Fe$_4$As$_4$F$_2$ at $\mu_0H_0$ = 11.999T and $T$ = 33K. The dark red arrow and pink arrow correspond to satellite peak of As1 site and As2 site respectively. Insets: central peaks after zooming in. Temperature dependence of NQR spectrums of As2 site (c) and As1 site (d).
\label{NMRNQRspecK}}
\end{figure}

Figure~\ref{NMRNQRspecK}(b) shows a typical NMR spectrum of KCa$_2$Fe$_4$As$_4$F$_2$. There are two types of As sites, as shown in Fig.~\ref{NMRNQRspecK} (a), namely As1 site close to K site and As2 close to Ca site~\cite{KCa2Fe4As4F2}. As mentioned above, there will be two center peaks for each $^{75}$As site, with low frequency peak and high frequency peak corresponding to $\theta$ = 42$^\circ$ and $\theta$ = 90$^\circ$ respectively. So the four peaks around the central transition of KCa$_2$Fe$_4$As$_4$F$_2$ are observed, which can be seen more clearly in the inset of Fig.~\ref{NMRNQRspecK}(b). We assign the inner two peaks coming from As1 site and outer two peaks from As2 site, as will be elaborated below.

\begin{figure}[htbp]
\includegraphics[width= 6 cm]{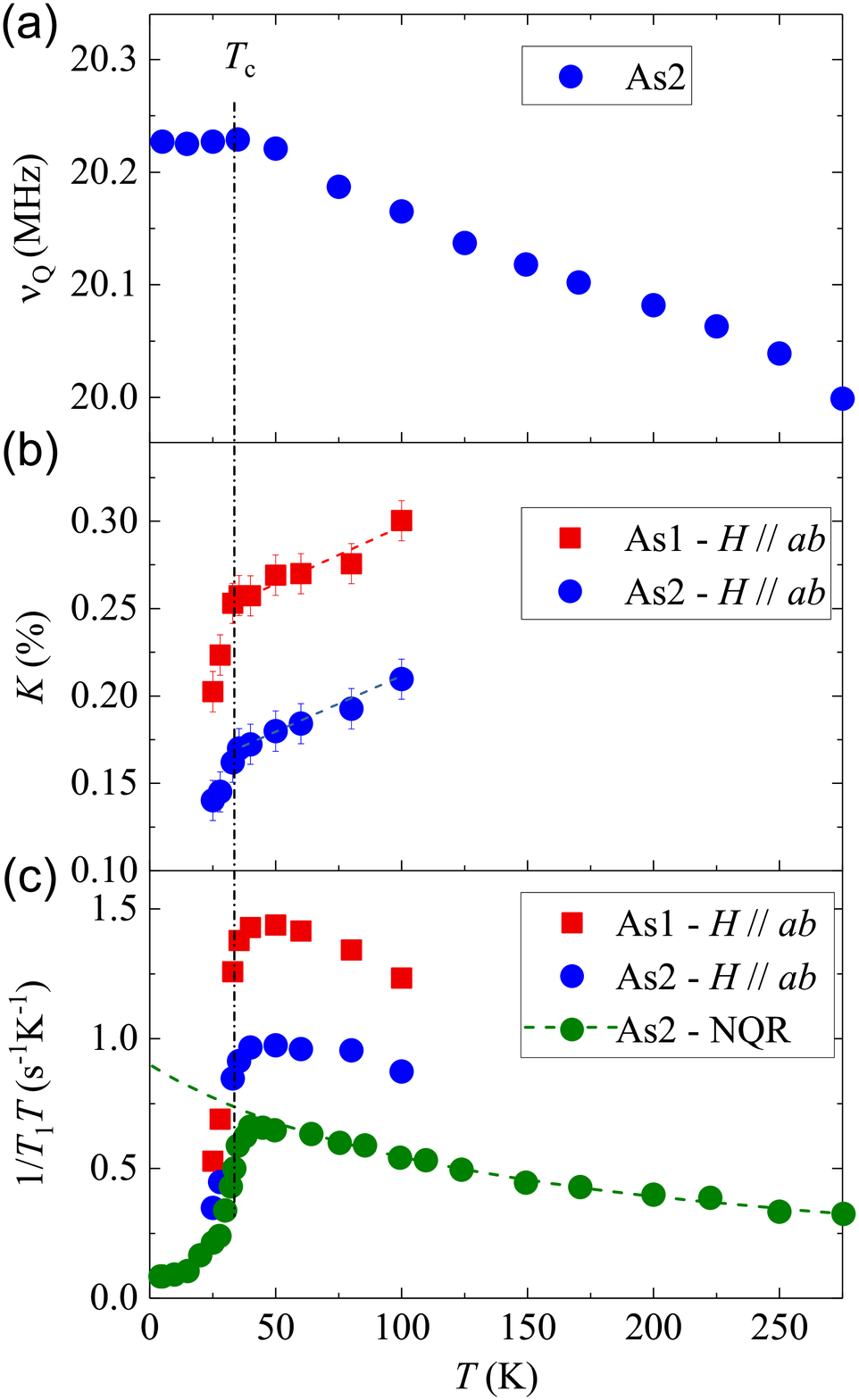}
\centering
\caption{(Color online) (a) Temperature dependence of $\nu_{\rm Q}$ of the As2 site. (b) Temperature dependence of Knight shift for two As sites.(c) Temperature dependence of 1/$T_1T$ measured at the two NMR peaks corresponding to $\theta$ = 90$^\circ$ at $H$ = 11.999T and As2 site at $H$ = 0 T. The olive dashed curve is a fitting by 2D antiferromagnetic fluctuations model. The black dashed-dotted line characterizes $T_{\rm c}$.
\label{T1TKvQ}}
\end{figure}

Figure~\ref{NMRNQRspecK}(c) shows the waterfall plot of NQR spectrums of As2 site at different temperatures. No splitting or broadening is seen in the NQR spectrums, indicating that magnetic order is absent in the studied compound. By using a Lorentz function to fit the spectrums, we deduced the temperature dependence of $\nu_{\rm Q}$, as summarized in Fig.~\ref{T1TKvQ}(a). The $\nu_{\rm Q}$ of As2 site increases from $T$ = 275K to 50K but saturates below 50K, as also seen in As2 site of CaKFe$_4$As$_4$~\cite{CaKFeAsNMR}. The $\nu_{\rm Q}$(100K) of As1 site is 10.9MHz and $\nu_{\rm Q}$(130K) of As2 site is 20.1MHz, close to $\nu_{\rm Q}$(100K) = 12.4MHz of KFe$_2$As$_2$~\cite{KFe2As2} and $\nu_{\rm Q}$(130K) = 19.2MHz of CaFeAsF~\cite{CaKFeAs}. we therefore assign the inner two peaks of central peaks corresponding to As1 site and the outer two peaks of central peaks corresponding to As2 site.  

\begin{figure}[t]
\includegraphics[width= 7.5 cm]{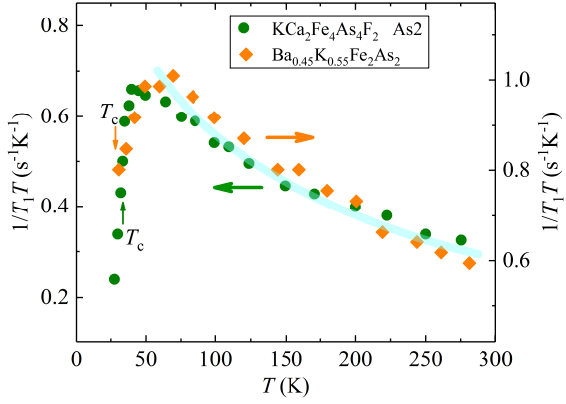}
\centering
\caption{(Color online) Comparison of 1/$T_1T$ for KCa$_2$Fe$_4$As$_4$F$_2$ and Ba$_{0.45}$K$_{0.55}$Fe$_2$As$_2$. For clarity, the starting point of right axis is 0.32s$^{-1}$K$^{-1}$ higher than that of left axis. The scale of two axis is the same.
\label{T1T12442vs122}}
\end{figure}

After subtracting the second order perturbation effect according to Eq.(2), the Knight shift for As1 site and As2 site, is obtained as shown in Fig.~\ref{T1TKvQ}(b). The Knight shift decreases from $T$ = 100K to $T_{\rm c}$, which is similar to most of iron based superconductors~\cite{LiFeAsK,BaKFeAsK,BaFeNiAsK,LaFeAsOFK,NaFeCoAsK}. The spin-lattice relaxation rate 1/$T_1$ was measured at two central peaks corresponding to $\theta$ = 90$^\circ$ which has a stronger intensity, and also at NQR peak of As2 sites. The results are presented in Fig.~\ref{T1TKvQ}(c). The magnitude of both $K$ and 1/$T_1T$ for As1 site is bigger than that of As2 site. This is likely due to different hyperfine coupling constant at the two sites. In iron-based superconductors, hyperfine coupling is determined by the overlap of the electron cloud between Fe and As, namely, by the distance between Fe and As. From the original data in ref~\cite{KCa2Fe4As4F2}, we obtain the distance between As and Fe plane \emph{h}(As-Fe) to be 1.405{\AA} and 1.436{\AA} for As1 site and As2 site, respectively. A smaller \emph{h}(As1-Fe) leads to a bigger hyperfine coupling constant, which can explain the larger $K$ and $1/T_1T$ for As1 site. The difference in the magnitude between 1/$T_1T$ obtained by NMR and NQR will be explained later. 1/$T_1T$ shows a monotonic increase as the temperature decreases from $T$ = 275K to 40K, indicating the existence of antiferromagnetic spin fluctuations in KCa$_2$Fe$_4$As$_4$F$_2$. We use a phenomenological 2D antiferromagnetic fluctuations model~\cite{Moriya,Moriya1} 1/$T_1T$$\propto$$a$+$C$/($T$ + $\varTheta$) to fit our data, where $C$/($T$ + $\varTheta$) is related to the low energy spin fluctuations and $a$ is due to other contributions. The fitting result is shown by olive dashed curve in Fig.~\ref{T1TKvQ}(c). We obtain $a$ = 0.015s$^{-1}$K$^{-1}$ and $\varTheta$ = 149K, respectively. Figure~\ref{T1T12442vs122} shows the comparison of 1/$T_1T$ between KCa$_2$Fe$_4$As$_4$F$_2$ and Ba$_{0.45}$K$_{0.55}$Fe$_2$As$_2$. After shifting the starting point of right axis upward by 0.32s$^{-1}$K$^{-1}$, we see that the 1/$T_1T$ of these two compounds are scaled very well. Thus it seems that the low energy spin fluctuations are quite similar for these two compounds. The dynamic susceptibility starts to decrease at $T$ $\approx$ 40 K (above $T_{\rm c}$), which is qualitatively similar to the features observed in cuperates~\cite{pseudogap}. This 'pseudogap' behavior was also observed in Ba$_{0.45}$K$_{0.55}$Fe$_2$As$_2$~\cite{BaKFeAsKgaplike} and over-doped La1111~\cite{LaFeAsOFgaplike,LaFeAsOFgaplike1}. The valence of Fe in KCa$_2$Fe$_4$As$_4$F$_2$ is +2.25, meaning that the equivalent doping level is 0.25 hole/Fe. This is close to the value in Ba$_{0.45}$K$_{0.55}$Fe$_2$As$_2$, where the doping level is 0.275 hole/Fe.

\begin{figure}[htbp]
\includegraphics[width= 7.5 cm]{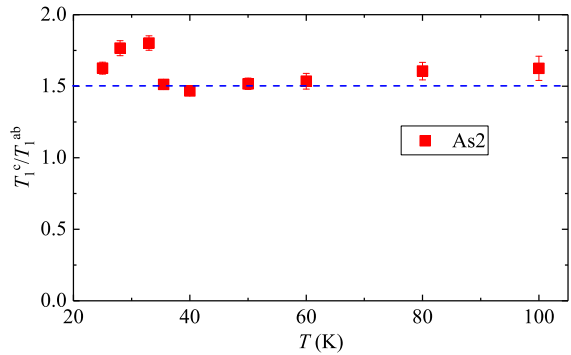}
\centering
\caption{(Color online) Temperature dependence of the ratio $T_{1}^{c}$/$T_{1}^{ab}$ for As2 sites.
\label{T1cvsT1ab}}
\end{figure}
To further study the nature of AFM spin fluctuations, we compare the 1/$T_1$ for $H_0$ parallel to $c$ direction and perpendicular to $c$ direction. In this compound, the principle axis of EFG is along $c$ direction. Therefore, $T_1$ measured in the NMR central peaks with $\theta$ = 90$^\circ$ corresponds to $T_{1}^{ab}$, while $T_1$ measured in the NQR peaks corresponds to $T_{1}^{c}$, as the principle axis of EFG is along the $c$ direction. Then we can obtain the anisotropy ratio of 1/$T_1T$, $R_{\rm AF}$ = $T_{1}^{c}$/$T_{1}^{ab}$. We find that it is only around 1.5 and nearly temperature-independent as shown in Fig.~\ref{T1cvsT1ab}. For the stripe order where the wave vector of spin fluctuations are [$\pi$, 0] and [0, $\pi$]~\cite{BaKFeAsK,T1abT1c}, $1/T_{1}^{ab}$ and $1/T_{1}^{c}$ can be expressed by

\begin{equation}
1/T_{1}^{ab} \propto A^2[\chi_{a}^{\prime \prime}\left(\omega_{0}, Q\right)+\chi_{b}^{\prime \prime}\left(\omega_{0}, Q\right)+\chi_{c}^{\prime \prime}\left(\omega_{0}, Q\right)]/2
\label{equ4}
\end{equation}

\begin{equation}
1/T_{1}^{c} \propto A^2\chi_{c}^{\prime \prime}\left(\omega_{0}, Q\right)
\label{equ5}
\end{equation}
where $A$ is hyperfine coupling constant and $\chi_{i}^{\prime \prime}(\omega_{0}, Q)$ is the imaginary part of the dynamic susceptibility along $i$ ($i$= $a$, $b$, $c$) direction at the measured angular frequency $\omega_{0}$. Therefore, one obtains
\begin{equation}
R_{\rm AF} = \frac{\chi_{a}^{\prime \prime}\left(\omega_{0}, Q\right)+\chi_{b}^{\prime \prime}\left(\omega_{0}, Q\right)}{2 \chi_{c}^{\prime \prime}\left(\omega_{0}, Q\right)}+\frac{1}{2}
\end{equation}
. If $\chi_{a}^{\prime \prime}\left(\omega_{0}, Q\right)$ = $\chi_{b}^{\prime \prime}\left(\omega_{0}, Q\right)$ = $\chi_{c}^{\prime \prime}\left(\omega_{0}, Q\right)$, the ratio $R_{\rm AF}$ will be equal to 1.5. Thus our observation suggests that the low energy spin fluctuations is isotropic in spin space.  This is in sharp contrast with the spin fluctuations of the optimally-doped Ba$_{0.68}$K$_{0.32}$Fe$_2$As$_2$, which is anisotropic~\cite{BaKFeAsK}. This result again indicates that the origin of the low energy spin fluctuations in KCa$_2$Fe$_4$As$_4$F$_2$ is different from the optimally-doped Ba$_{0.68}$K$_{0.32}$Fe$_2$As$_2$. The anisotropic spin fluctuations in Ba$_{0.68}$K$_{0.32}$Fe$_2$As$_2$ was ascribed to a spin-orbit coupling (SOC) which was estimated to be 10$\sim$20 meV. Our result therefore suggests that the SOC is smaller in  KCa$_2$Fe$_4$As$_4$F$_2$. Rather, the stoichiometric compound KCa$_2$Fe$_4$As$_4$F$_2$ provides a unique platform for studying the overdoped region of iron-based superconductors. Eq.~\ref{equ4} and ~\ref{equ5} also explain why $1/T_{1}T$ obtained by NQR is smaller than that by NMR, as in the former case the quantized axis is along the \emph{c}-axis.

\begin{figure}[htbp]
\includegraphics[width= 5.5 cm]{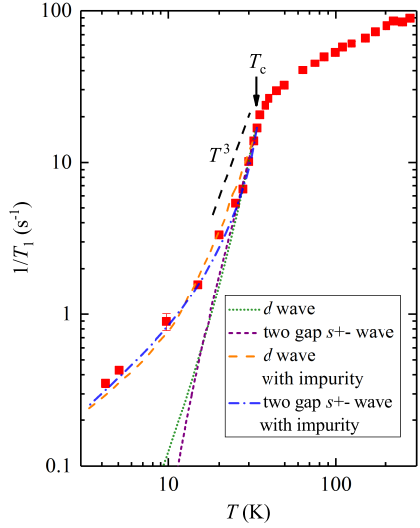}
\centering
\caption{(Color online) Temperature dependence of 1/$T_1$ in KCa$_2$Fe$_4$As$_4$F$_2$. The black dashed line is a guide to the eyes, indicating a $T^3$ variation of 1/$T_1$. The green dotted, purple short-dashed, blue dashed-dotted and orange dashed curve shows the fitting results of $d$ wave, two-gap $s$+- wave, $d$ wave with impurity scattering and two-gap $s$+- wave with impurity scattering, respectively.
\label{12442SC}}
\end{figure}
Finally, we discuss the property in the superconducting state. Figure~\ref{12442SC} shows the temperature dependence of 1/$T_1$. Below $T_{\rm c}$, no coherence peak appears and $T_1$ decreases more rapidly than $T^3$. Below $T$ = 15K, 1/$T_1$ starts to be proportional to $T$, indicating the existence of  strong impurity scattering in this sample~\cite{dwavedeviateT3,dwavewithimpurity}. We further simulate our results by assuming different gap symmetries as shown in Fig.~\ref{12442SC}. In the past decade of researches on iron-pnictides, multiple gaps have been found~\cite{splusminus,splusminus1,caotwogap}. In fact, $s$+- wave gap symmetry in which  gap sign reverses between hole fermi pocket and electron fermi pocket can account for the hump behavior of  1/$T_1$ in  various compounds~\cite{LiFeAsK,BaKFeAsK,LaFeAsOFgaplike} . Due to strong impurity scattering, however, the hump feature in 1/$T_1$ is not visible in the present case. We tried to fit our data with various models.  A simple $d$-wave model with $\Delta^d$ = 2.5$k_BT_{\rm c}$ or a two-band (two gap) $s$+- wave with $\Delta^s_1$ = 2$k_BT_{\rm c}$, $\Delta^s_2$ = 3.75$k_BT_{\rm c}$ and equal weight for the two band deviate from our data severely at low temperatures. Following the $T_1$ calculation method of $d$-wave with impurity in literature~\cite{dwavewithimpurity}, we found that the parameter $\Delta^d$ = 2.5$k_BT_{\rm c}$, $\eta$ = 0.064$\Delta^d$ can fit our data well. On the other hand, two-gap $s^{+-}$-wave with impurity scattering can also account for our data. In the $s^{+-}$-wave model~\cite{splusminusfitting}, 1/$T_1$ is expressed as:
\begin{equation}
\begin{aligned}
\frac{1}{T_{1}} &\sim-T \int_{0}^{\infty} d\omega \frac{\partial f(\omega)}{\partial \omega}\left(W_{\rm GG}+W_{\rm FF}\right)\\
W_{\rm GG}&=\left[\sum_{a=h ,e} N_{a}(0)\left\langle\operatorname{Re} \frac{\omega}{\sqrt{\omega^{2}-\Delta_{a}^{2}(k)}}\right\rangle_{k}\right]^{2}\\
W_{\rm FF}&=\left[\sum_{a=h,e} N_{a}(0)\left\langle\operatorname{Re} \frac{\Delta_{a}(k)}{\sqrt{\omega^{2}-\Delta_{a}^{2}(k)}}\right\rangle_{k}\right]^{2}.
\end{aligned}
\end{equation}
Where $N_{a}$ is DOS in hole or electron fermi surface. We define $\alpha$ = $\frac{N_{e}}{N_{e}+N_{h}}$. If there exists impurity scattering, $\omega$ will be replaced by $\omega + i\eta$. Using this model, we found that the parameters $\Delta^s_1$ = 2$k_BT_{\rm c}$, $\Delta^s_2$ = 3.75$k_BT_{\rm c}$, $\alpha$ = 0.5, $\eta$ = 0.21$\Delta^s_1$ reproduce our data well. In order to distinguish $d$-wave and $s^{+-}$-wave, more measurements on a single crystal are required.

\subsection{A$_2$Cr$_3$As$_3$}
\begin{figure}[htbp]
\includegraphics[width= 10 cm]{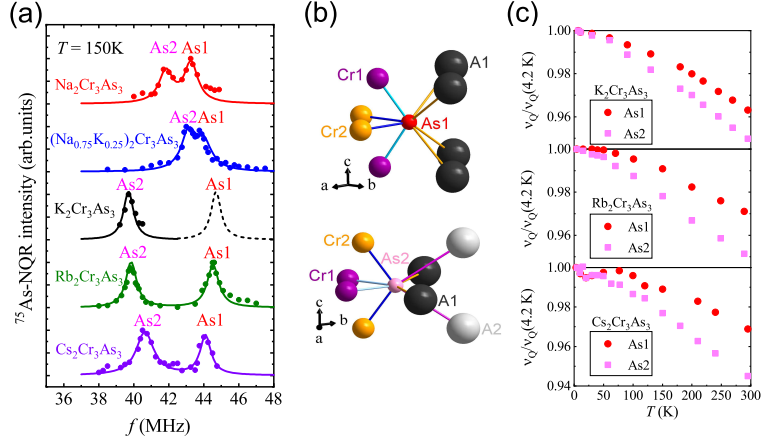}
\centering
\caption{(Color online) (a) $^{75}$As NQR spectra of A$_2$Cr$_3$As$_3$(A = Na, Na$_{0.75}$K$_{0.25}$, K, Rb, Cs) measured at $T$ = 150 K. The two peaks correspond to two As sites, namely As1 site and As2 site. The data for the right peak of K$_2$Cr$_3$As$_3$ was referred from Ref.~\cite{imai}. (b) Surrounding environments of As1 site and As2 site, respectively. (c) Temperature dependence of $\nu_{\rm Q}$ normalized by its value at 4.2K. The original data of K$_2$Cr$_3$As$_3$ and Cs$_2$Cr$_3$As$_3$ were referred from Ref.~\cite{imai,imaiCsCrAs}.
\label{specofA2Cr3As3}}
\end{figure}

\begin{figure}[htbp]
\includegraphics[width= 6 cm]{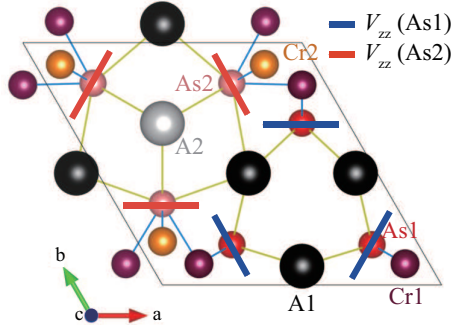}
\centering
\caption{(Color online) EFG principle axis ($V_{ZZ}$) for K$_2$Cr$_3$As$_3$, which are shown by the blue and red sticks for As1 site and As2 site, respectively.
\label{K2Cr3As3Vzz}}
\end{figure}

\begin{table}[htbp]
\centering
\caption{Comparison of experimental and theoretical results of $\nu_{\rm Q}$ for K$_2$Cr$_3$As$_3$. The experimental data were obtained at 300 K. In the EFG calculation, the lattice constant at 300K which was referred from Ref.~\cite{K2Cr3As3} was used. The units of $V_{\alpha\alpha}$ = $\frac{ \partial{V^2}}{\partial {\alpha^2}}$ ($\alpha$ = $X$, $Y$, $Z$) and $\nu_{\rm Q}$ are 10$^{19}$V/m$^2$ and MHz, respectively.}
\label{table1}
\begin{tabular}{|c|cccc|c|}
\hline  
site& $V_{XX}$& $V_{YY}$& $V_{ZZ}$& $\nu_{\rm Q}$& $\nu_{\rm Q}$(exp)\\
\hline  
As1&  -526.35&  -570.91& 1097.27& 41.64 & 43.6 \\

As2&  -524.51& -528.48& 1052.98& 39.94 & 38.4 \\

\hline 
\end{tabular}\\
\end{table}
Next, we turn to A$_2$Cr$_3$As$_3$ (A = Na, Na$_{0.75}$K$_{0.25}$, K, Rb, Cs). Figure.~\ref{specofA2Cr3As3}(a) shows the NQR spectra for A$_2$Cr$_3$As$_3$ at $T = $150 K, the five samples commonly show two peaks originating from two inequivalent As sites. 
In the previous reports, the site assignment was not performed. We assign the left peak to As2 site and the other peak to As1 site based on the following two facts. Firstly, neutron diffraction measurements found a displacement of the K2 sites towards CrAs tube with decreasing temperature in K$_2$Cr$_3$As$_3$~\cite{structureins}. As shown in Fig.~\ref{specofA2Cr3As3}(b), the As1 site has four nearest A1 neighbors, while the As2 has two nearest A1 neighbors and two nearest A2 neighbors. Therefore, the displacement of the A2 sites makes a stronger variation of the EFG for As2 sites, leading to a stronger temperature variation of $\nu_{\rm Q}$ for As2 site than As1 site. Indeed, the change of $\nu_{\rm Q}$ defined as $\delta$$\nu_{\rm Q}$ = $\delta$$\nu_{\rm Q}(T=4.2 \rm K)$ - $\delta$$\nu_{\rm Q}(T=300 \rm K)$, is bigger for the left peak than the right peak~\cite{imai,imaiCsCrAs,yang}. It can be seen more clearly in Fig.~\ref{specofA2Cr3As3}(c), where we normalize the original $\nu_{\rm Q}$ data by $\nu_{\rm Q}(T=4.2\rm K)$. Secondly, we calculate the EFG of two As sites for K$_2$Cr$_3$As$_3$.  The calculation is based on HiLAPW which is a extension of FLAPW~\cite{HILAPW}. We use Generalized-Gradient Approximation(GGA) as exchange correlation function, and spin orbital coupling(SOC) are included. The nuclear quadrupole moment $Q$ =  3.141$\times$10$^{-29}$m$^2$ was used for nucleus $^{75}$As~\cite{Q}. The inputting lattice constant is from Ref~\cite{K2Cr3As3} at $T$ = 300K. The calculational result is shown in Table.~\ref{table1}. The EFG tensors  are defined as $V_{\alpha\alpha}$ = $\frac{ \partial{V^2}}{\partial {\alpha^2}}$ ($\alpha$ = $X$, $Y$, $Z$), where $V$ is electric potential. In case of $\eta$ = $\frac{V_{XX}-V_{YY}}{V_{ZZ}}$ $\neq$ 0, the $\nu_{\rm Q}$ is defined as $\frac{3e^2qQ\sqrt{1+\eta^2/3}}{2I(2I-1)h}$, where EFG $eq$ = $V_{ZZ}$ and $V_{ZZ}$ is the principle-axis value. As shown in Fig.~\ref{K2Cr3As3Vzz}, the principle axis of A$_2$Cr$_3$As$_3$ is in the $ab$ plane. The resulting $\nu_{\rm Q}$ = 41.64MHz of As1 site is bigger than $\nu_{\rm Q}$ = 39.94MHz of As2 site, which supports the above mentioned site assignment. The site assignment will help further study the physical properties of two As sites by NMR, which is vitally important to identify the pairing symmetry.

Before closing, we note that the three transition-metal arsenides reported here show quite different normal-state properties and the superconducting gap symmetry in these compounds is also different. LaRu$_2$As$_2$ shows superconductivity with a full gap, whose origin maybe electron-phonon coupling~\cite{LaRuAsDFT,LaRuAsab}. On the other hand, unconventional superconductivity caused by AFM spin fluctuations in KCa$_2$Fe$_4$As$_4$F$_2$ and FM spin fluctuation in A$_2$Cr$_3$As$_3$~\cite{yang,FMQCP}. Future issues include clarifying whether spin triplet pairing is realized by the FM spin fluctuation in A$_2$Cr$_3$As$_3$.

\section{Summary}  
We have performed NMR and NQR measurement in three types of transition metal arsenides, LaRu$_2$As$_2$, KCa$_2$Fe$_4$As$_4$F$_2$, and A$_2$Cr$_3$As$_3$. In LaRu$_2$As$_2$, different from Fe-based superconductors with the same crystal structure, a coherence peak in the temperature dependence of 1/$T_1$ appears just below $T_{\rm c}$, indicating that the superconducting gap is fully opened. In double Fe$_2$As$_2$ layers compound KCa$_2$Fe$_4$As$_4$F$_2$, the strength of antiferromagnetic spin fluctuations are found to be similar to that in Ba$_{0.45}$K$_{0.55}$Fe$_2$As$_2$, indicating that the stoichiometric compound KCa$_2$Fe$_4$As$_4$F$_2$ is in the moderately hole-overdoped region. In fact, the anisotropy of 1/$T_1$, $R_{\rm AF}$ = $T_{1}^{c}$/$T_{1}^{ab}$ is only 1.5, implying that the spin fluctuations are isotropic,  which is in sharp contrast to the nearly optimally-doped Ba$_{0.68}$K$_{0.32}$Fe$_2$As$_2$. For A$_2$Cr$_3$As$_3$, we identified the one-to-one correspondence between NQR peaks and As sites. Our research revealed a wide variety of the physical properties of transition metal arsenides.

\textbf{Acknowledgments} We thank Y. G. Shi, C. J. Yi, Q. G. Mu, Z. T. Tang, and T. Liu for early collaborations. Project supported by the National Natural Science Foundation of China (No. 11674377, No. 11634015, No. 11974405), the National Key R$\&$D Program of China (No. 2017YFA0302904 and No. 2016YFA0300502),J. Y. also acknowledges support by the Youth Innovation Promotion Association of CAS.

\end{document}